\documentclass{aa} 
\usepackage{graphicx}
\usepackage{txfonts}
\begin{document}
\title{Obscured clusters.\,II. GLIMPSE-C02 -- A new\\ metal rich
globular cluster in the Milky Way\thanks{Based on observations
collected with the ESO New Technology Telescope, observing program
77.D-0089.}\fnmsep\,\thanks{Table with photometry is only available 
in electronic form at the CDS via anonymous ftp to cdsarc.u-strasbg.fr 
(130.79.128.5) or via http://cdsweb.u-strasbg.fr/cgi-bin/qcat?J/A+A/
 }}

\subtitle{}

\author{R. Kurtev\inst{1}\fnmsep\,\thanks{``Centro de Astrof\'isica
de Valpara\'so''. Visiting astronomer at the ESO La Silla Paranal
Observatory.}
\and
V.D. Ivanov\inst{2}
\and
J. Borissova\inst{1}$^{,\,\star\star\star}$
\and
S. Ortolani\inst{3}
}

\offprints{R. Kurtev}

\institute{Departamento de F\'isica y Astronom\'ia,
Facultad de Ciencias, Universidad de Valpara\'{\i}so,
Ave. Gran Breta\~na 1111, Playa Ancha, Casilla 53,
Valpara\'iso, Chile \\
\email{radostin.kurtev@uv.cl; jura.borissova@uv.cl}
\and
European Southern Observatory, Ave. Alonso de
Cordova 3107, Casilla 19, Santiago 19001, Chile\\
\email{vivanov@eso.org}
\and
Universit\'a di Padova, Dipartimento di Astronomia,
Vicolo dell'Osservatorio 5, I-35122 Padova, Italy
\email{sergio.ortolani@unipd.it}
}

\date{}

\abstract{The estimated total number of Milky Way globulars is 160$\pm$20. 
The question of whether there are any more
undiscovered globular clusters in the Milky Way is particularly
relevant with advances in near and mid-IR instrumentation.}
{This investigation is a part of a long-term project to search
the inner Milky Way for hidden star clusters and to study them
in detail. GLIMPSE-C02 (G02) is one of these objects, situated
near the Galactic plane ($l$=$14\fdg129$, $b$=$-0\fdg644$).}
{Our analysis is based on SOFI/NTT $JHK_S$ imaging and
low resolution (R$\sim$1400) spectroscopy  of three bright cluster
red giants in the $K$ atmospheric window. We derived the metal
abundance by analysis of these spectra and from the slope of the RGB.}
{The cluster is deeply embedded in dust and undergoes a mean reddening
of $A_{V}$\,$\sim$\,24.8$\pm$3\,mag. The distance to the object is
D=4.6$\pm$0.7\,kpc. The metal abundance of G02 is
[Fe/H]$_{\rm H96}$=$-$0.33$\pm$0.14 and [Fe/H]$_{\rm CG}$=$-$0.16$\pm$0.12 
using different scales. The best 
fit to the radial surface brightness profile with a single-mass 
King's model yields a core radius $\rm r_c$=0.70 arcmin (0.9\,pc), 
tidal radius $\rm r_t$=15 arcmin (20\,pc), and central concentration 
c=1.33.}
{We demonstrate that G02 is new Milky Way globular cluster, among the 
most metal rich globular clusters in the Galaxy. The object is physically located 
at the inner edge of the thin disk and the transition region with 
the bulge, and also falls in the zone of the ``missing'' 
globulars toward the central region of the Milky Way.}


\maketitle

\section{Introduction}

Star clusters provide us with unique conditions to
investigate various aspects of stellar astrophysics under
tightly ``controlled'' conditions -- the clusters are samples of
stars with similar ages, metallicities and distances. The
Galactic globular clusters (GCs) can be used to collect
information about the formation and early evolution of the Milky
Way.

The large area infrared (IR) surveys (i.e. 2MASS, Skrutskie et
al. \cite{skr06}) have discovered a number of new clusters,
hidden by the dust extinction in the plane of the Milky Way.
These objects usually suffer A$_V$$\geq$10-20\,mag of extinction,
making them invisible in the optical wavebands.
The vast majority of them appear to be a few million years old
(Ivanov et al. \cite{iva02}, \cite{iva05}; Borissova et al.
\cite{bor03}, \cite{bor05}, \cite{bor06}; Kurtev et al. \cite{kurt07}) but a
few have proved to be analogues of ``classical'' globular clusters
(Hurt et al. \cite{hur00}, Ortolani, Bica \& Barbuy \cite{ort00},
Kobulnicky et al. \cite{kob05}, Carraro \cite{car05}, Froebrich,
Meusinger \& Scholz \cite{froe07}). There are probably $\sim$10
``missing'' globulars in the central region of the Milky Way,
based on the asymmetry of the GC distribution (Ivanov, Kurtev \&
Borissova \cite{iva05a}).

\begin{figure}
\centering
\includegraphics[width=\columnwidth]{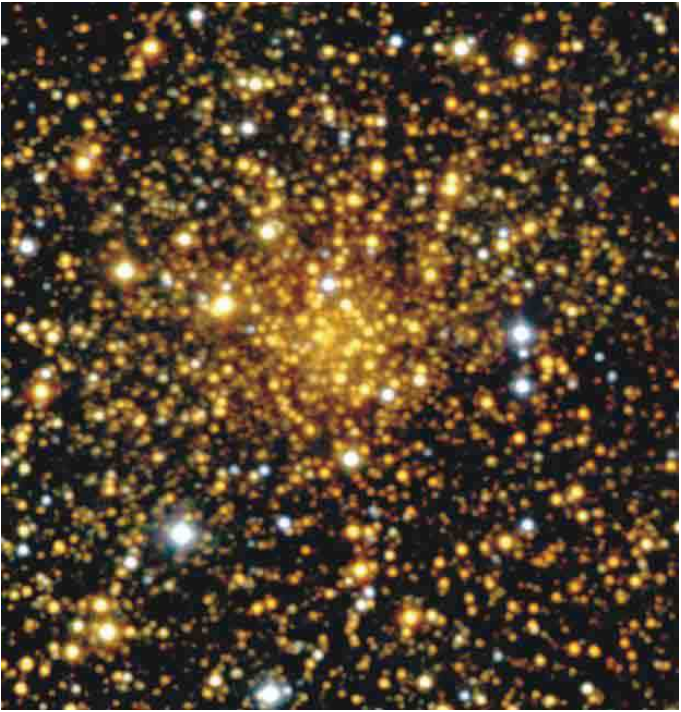}
\caption{Pseudo-true color images of G02. It is composed
from our SofI/NTT $J$ (blue), $H$ (green) and $K_S$ (red) data. The
field of view is $\sim$3$\times$3\,acmin centered at the cluster
with coordinates: $\alpha(2000)=18^{\rm h}\,18^{\rm m}\,30\fs5$ 
and $\delta(2000)=-16\degr\,58'\,38''$. North is up and East is 
to the left.}
\label{fig1}
\end{figure}

{\it Spitzer Space Telescope} Galactic Legacy Infrared Mid-Plane
Survey Extraordinaire (GLIMPSE, Benjamin et al. \cite{ben03})
offers an excellent opportunity to carry out an even deeper census of
such objects than it is possible in the near-IR because of the
lower extinction at longer wavelengths. A comprehensive search
for clusters using the point source catalog of GLIMPSE found 92
candidates (Mercer et al. \cite{mer05}). As a part of
our long-term project to find and characterize new Milky Way
clusters we studied some of them using deep near-IR imaging and
low resolution IR spectroscopy. In the course of this study we
concluded that the candidate Nr. 3 in their list
(Fig.\,\ref{fig1}) is a new Galactic metal rich globular
cluster. Here we present the evidence and we report its
properties -- metallicity, extinction and distance. 
We will refer to the new cluster as GLIMPSE-C02 (G02), for 
consistency with the designation of Kobulnicky et al. (\cite{kob05}).

\section{Observations and data reduction}

All observations were obtained with SofI/NTT (Son of ISAAC;
Moorwood, Cuby \& Lidman \cite{mor98}) on Apr 15, 2006. The
instrument was equipped with a
Hawaii HgCdTe 1024$\times$1024 detector, with a pixel scale of
0.288\,arcsec\,px$^{-1}$. For the spectroscopy we used a
1\,arcsec slit and the medium-resolution grism, yielding a
resolution of R$\sim$1320 at 2.2\,$\mu$m. The seeing for all
observations was 1-1.5\,arcsec and the sky was photometric.

We collected a total of 16\,min integration in each of the $JHK_S$
filters, split into 16 images, jittering within a 3\,arcmin square
box to ensure that there is minimum overlapping of the cluster
position. Each individual image was the average of
3$\times$20\,sec frames in $J$, 6$\times$10\,sec frames in $H$,
and 10$\times$6\,sec frames in $K_S$. The data reduction
included flat fielding, sky subtraction, alignment and
combination of the individual images. The stellar photometry was
carried on the final images with {\sc ALLSTAR} in
{\sc DAOPHOT\,II} (Stetson \cite{ste93}). The typical photometric
errors vary from 0.01\,mag for stars with $K_S$$\sim$10\,mag to
0.10\,mag for $K_S$$\sim$18\,mag and 0.15\,mag for
$K_S$$\sim$19\,mag. The photometric calibration was performed by
comparing our instrumental magnitudes with the 2\,MASS
measurements of about 1200 stars, covering the color range
0.0$\leq$$J$$-$$K_S$$\leq$6.0\,mag and magnitude range
10.0$\leq$$K_S$$\leq$15.0\,mag. The final photometry list contains
equatorial coordinates and $JHK_S$ magnitudes of 7623 stars with
photometric errors less than 0.15\,mag. Artificial star tests
show that the 80\% completeness limit of the photometry is at
$J$=18.9 and $K_S$=17.3\,mag.

The spectra cover the $\sim$2.00-2.35\,$\mu$m region. Fortuitously,
the slit could be placed to contain 12 stars. The telescope was
nodded along the slit between the exposures to simultaneously observe 
the targets and clear sky. In total, we obtained 8 images of
300\,sec. First, we flat fielded them and removed the sky emission
by subtracting from each field the images from each nodding pair.
Next, we extracted 1-dimensional spectra with the {\sc IRAF}\footnote{IRAF 
is distributed by the National Optical Astronomy Observatory, which is 
operated by the Association of Universities for Research in Astronomy, 
Inc., under cooperative agreement with the National Science Foundation 
(NSF).} task {\sc APALL}, wavelength calibrated them with the NeXe lamp spectra
(extracted at the location of the science target spectra), and
combined them into final 1-dimensional spectra. Finally, we divided
them by the spectra we took from the solar near-analog HIP\,59642
(HD\,106290) of type G1V, and multiplied them by a solar spectrum to
remove the artificial emission lines due to the intrinsic absorption
features in the spectra of the standard (see Maiolino, Rieke \&
Rieke \cite{mai96}).

\section{Properties of G02}

\subsection{Extinction and distance}

\begin{figure}
\resizebox{\hsize}{!}{\includegraphics{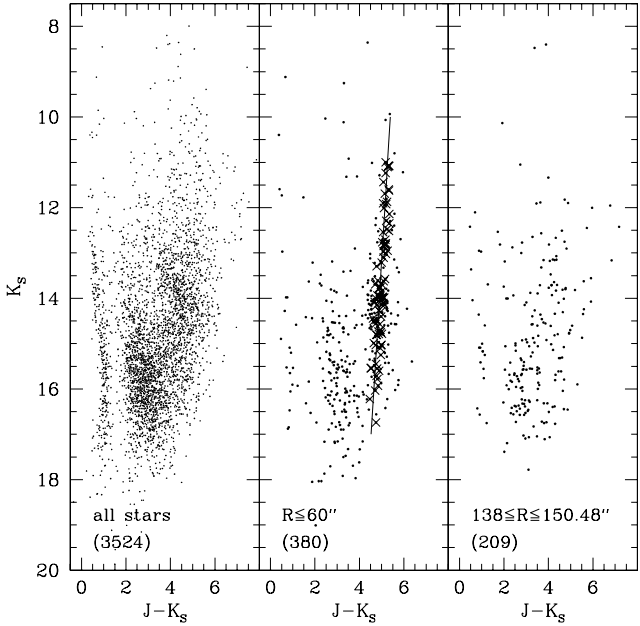}}
\caption{Near-IR color-magnitude diagram for all stars on our
images (left), and the stars in two regions with identical
areas -- a circle with radius R=60 arcsec centered at the
cluster center (middle), and an annulus centered at the cluster
with an inner radius of 138 arcsec and an outer radius of 150.48 arsec
 (right). The numbers of stars
plotted in each panel are given in brackets. The solid line in the 
middle panel is
the best fit to the RGB and the crosses mark the stars used to
derive the fit in one Monte-Carlo realization (see
Section~\ref{Section_MC} for details).}
\label{fig2}
\end{figure}

The color-magnitude diagram (CMD) of G02 is plotted in
Fig.~\ref{fig2}. The left panel contains all stars with $J$ and
$K_S$ photometry. It is clearly dominated by field stars. Two
main sequences (MS) are evident: a nearby, bright one at
$J$$-$$K_S$$\sim$0-1\,mag and $K_S$$\sim$8.5-17.5\,mag, and a
distant, reddened one at $J$$-$$K_S$$\sim$2-4\,mag and
$K_S$$\sim$12-17.5\,mag. There is an indication of a red clump
sequence, starting at $J$$-$$K_S$$\sim$3\,mag, $K_S$$\sim$13.3\,mag
and extending redwards toward fainter magnitudes due to extinction.
The cluster red giant branch (RGB) is distinctly visible at
$J$$-$$K_S$$\sim$4.5-5.5\,mag in the middle panel where we
intentionally selected a relatively small region with radius R=60 arcsec
near the center of G02 to reach good cluster-to-field contrast. Finally, 
the ``field'' annulus on the right panel, with an inner radius of 138 
arcsec and an outer radius of 150.48 arsec and identical area, contains 
a mixture of populations and it probably still includes some cluster 
members but we refrain from using a more distant comparison area because 
of the clumpy dust absorption that might severely affect the number of 
field stars and compromise the background subtraction.

\begin{figure}
\resizebox{\hsize}{!}{\includegraphics{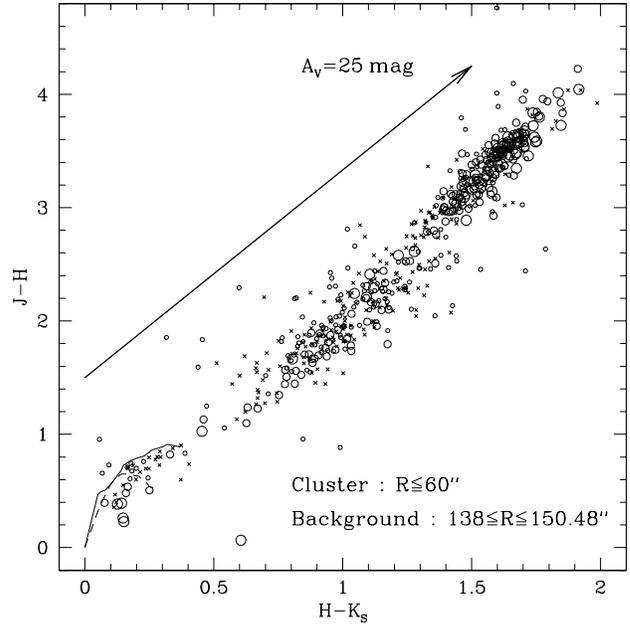}}
\caption{Near-IR color-color diagram for stars in two
regions in the vicinity of G02, with identical areas --
a circle with radius R=60\,arcsec around the cluster center
(open circles; the size is proportional to the apparent
brightness of the star), and an annulus around the cluster
with inner radius of 138\,arcsec and an outer radius of 150.48 
arsec (crosses). A vector
representing visual extinction of A$_V$=25\,mag is plotted,
using the reddening law of Rieke \& Lebofsky (\cite{rie85}).
The face-value colors of RGB (solid line) and MS stars
(dashed line) from Frogel et al. (\cite{fro78}) are shown.}
\label{fig3}
\end{figure}

The color-color diagram of the cluster field is shown in
Fig.~\ref{fig3}. A comparison of the locus, occupied by
cluster stars, with the unreddened RGB and MS sequences of
Frogel et al. (\cite{fro78}) yields extinction toward G02
of A$_V$$\sim$24.5$\pm$3\,mag corresponding to a K-band
extinction of A$_K$$\sim$2.75$\pm$0.3\,mag. Some of the
extinction spread might be contributed to differential
extinction across the face of the clusters, contamination
from extended emission and background sources. Here and
throughout the rest of this paper we use the reddening law
of Rieke \& Lebofsky (\cite{rie85}). Interestingly, the
reddening line appears to deviate slightly from the
sequence of the reddened stars which may indicate anomalous
dust properties in the direction toward G02.

The cluster luminosity function (LF) is shown in
Fig.~\ref{fig3}. The presence of red giant clump stars is
evident and a Gausian fit of the bump in the LF gives a
mean value of $K_S$=14.4$\pm$0.15\,mag. The clump has an
absolute magnitude of $M_K$=$-$1.83$\pm$0.03\,mag (Alves
\cite{alv00}, Eqn.~3), yielding a distance modulus of
$(m$$-$$M)_0$$\sim$13.45$\pm$0.3\,mag. Here we used
[Fe/H]=$-$0.33 (see Section\ref{Section_MC}).

The distance to the cluster can also be measured from the
RGB tip brightness (Ivanov \& Borissova \cite{iva02}). This
method is hampered by the small number of stars at the upper
end of the RGB in GCs. Nevertheless, we applied this test as
a consistency check. Assuming [Fe/H]=$-$0.33 we obtain an
absolute K-band magnitude for the tip of $M_K$=$-$6.8\,mag
(Ivanov \& Borissova \cite{iva02}). The apparent magnitude
of the tip is $K_S$$\sim$9.5\,mag, yielding
$(m$$-$$M)_0$$\sim$13.52\,mag, in excellent agreement with our
previous estimate.

We calculated the reddening and distance using the Padova isochrones 
(Girardi et al. \cite{girar00})
interpolated for metallicity [Fe/H]=$-$0.33.
The RGB color at the level of the HB is $J$-$K$=4.75. The true color, 
obtained from the isochrones, is 0.7 which yields $E$($J$-$K$)=4.05 
mag, corresponding to $A_V$=4.05x6.2=25.1, consistent 
with our previous discussion. The yielded distance modulus using 
$M_K$=$-$1.55 (from Padova isochrones) is $\sim$12.85, 
close to both upper values. 

Finally, we obtain the mean values of A$_V$$\sim$24.8$\pm$3\,mag, $(m$$-$$M)_0$$\sim$13.3$\pm$0.3\,mag, and D=4.6$\pm$0.7\,kpc for 
the reddening, the true distance modulus, and the distance to 
the cluster.

\begin{figure}
\resizebox{\hsize}{!}{\includegraphics{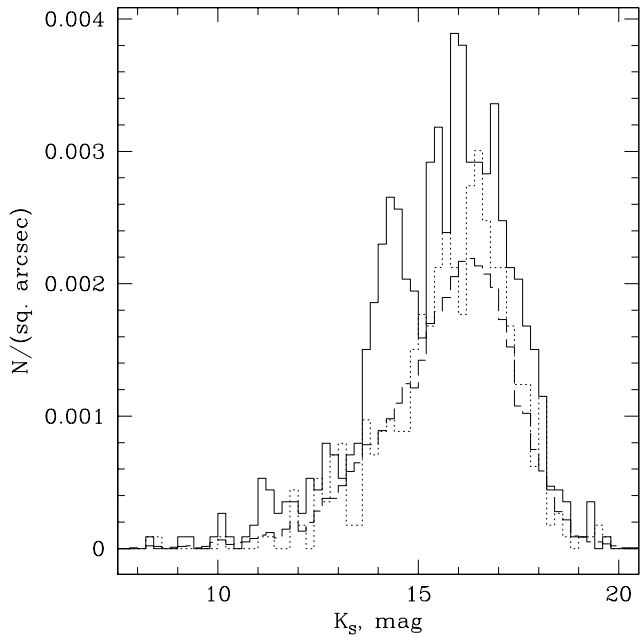}}
\caption{Normalized near-IR luminosity function, in stars
per square arc second, for stars within 60\,arcsec from the
center of the cluster (solid line), stars in an annulus
with an inner radius of 138 arcsec and an outer radius of 150.48 
arsec and the same area as the
cluster region (dotted line), and all stars in the field of
G02 (long-dashed line). The last function is normalized to 
the whole image area of 7.4\,$\times$\,7.4 arcmin.}
\label{fig4}
\end{figure}

\subsection{Metal abundance\label{Section_MC}}

We derived the metal abundance of the cluster using two
independent techniques. First, from the spectra of three
cluster red giants (listed in Table~\ref{tab1}), we applied 
the method of Frogel et al. (\cite{fro01}), which is based on the
behavior of the K-band spectral indices, IR magnitudes,
and colors. The individual reddening and true color of each star
are obtained by moving the star along the reddening vector until 
it crosses the luminosity class III sequence taken from Frogel 
et al. (\cite{fro78}).
All these stars are located near the cluster center, and
two of them (\#1 and \#3) lie directly on the cluster RGB, 
making them highly probable members. The last one (star \#2) 
suffers from less reddening ($\Delta$$A_K$$\sim$0.8\,mag), 
but the observed differential reddening variations could account 
for this. Indeed, the width of the cluster star locus on the color-magnitude 
and color-color diagram suggests reddening variations across 
the face of the cluster of the order of $\Delta A_K$$\sim$0.3 mag.
Therefore, the reddening of star \#2 is identical to that of the 
other two red giants with spectra, within 3$\sigma$ and differs in 
less than 2$\sigma$ from the mean reddening toward the cluster. 
Furthermore, its angular separation of only 12 arcsec from the
cluster center strengthens the possibility that this star is 
a cluster member.

\begin{figure}
\resizebox{\hsize}{!}{\includegraphics{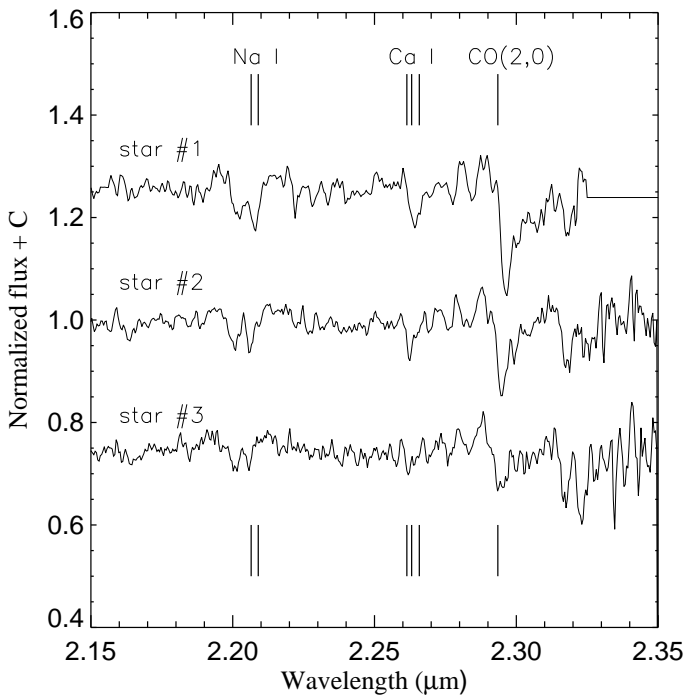}}
\caption{K-band spectra of three stars in G02. The position of the Na
doublet, the Ca triplet, and the first band of CO are indicated.
}
\label{fig5}
\end{figure}

The spectra are plotted in Fig.~\ref{fig5}. We measured
the equivalent widths of the Na\,2.20\,$\mu$m,
Ca\,2.26\,$\mu$m, and CO\,2.3\,$\mu$m features, and applied
the calibration (eqn. 4 of Frogel et al. \cite{fro01}) that takes into 
account also the stellar color ($J$-$K$)$_0$ and absolute magnitude $M_K$. 
The results are listed in Table~\ref{tab1}. 
Averaging the three estimates from the spectroscopy we obtain 
[Fe/H]=$-$0.29$\pm$0.10, in the metallicity scale of Zinn 
(as implemented in Harris \cite{har96}). Omitting star \#2, we obtain [Fe/H]=$-$0.34$\pm$0.10. The remaining nine stars with spectra 
were not suitable for metallicity measurements either because 
they were too faint or because they were early-type field stars.

\begin{table}
\caption{Spectroscopic measurements of red giant stars in G02.}

\medskip
\tabcolsep=5.1pt
\small
\begin{center}
\begin{tabular}{lccccccc}
\hline\hline
Star & \multicolumn{3}{c}{EW, \AA} & ($J$-$K$)$_0$ & $A_K$ & $M_K$ & [Fe/H]\\
& Na & Ca & CO & mag & mag & mag & \\
\hline
\#1 & 4.78 & 3.62 & 15.44 & 0.826 & 3.002 & $-$4.828 & $-$0.29 \\
\#2 & 5.88 & 2.35 & 12.48 & 0.601 & 2.188 & $-$4.324 & $-$0.20 \\
\#3 & 4.89 & 1.57 & 10.33 & 0.722 & 2.970 & $-$4.616 & $-$0.39 \\
\hline
\end{tabular}
\end{center}
\label{tab1}
\begin{list}{}{}
\item[] The table contains 
the equivalent widths of Na, Ca and CO lines, true color ($J$-$K$)$_0$, individual reddening  ($A_K$), absolute $K$ magnitude ($M_K$), and metallicity in the scale of Zinn.
In the calculation of the absolute magnitude we used the mean distance modulus 
to the cluster: $(m$$-$$M)_0$$\sim$13.3.
\end{list}
\end{table}

Next, we derived the abundances from the RGB slope. This
method was pioneered by Da Costa \& Armandroff (\cite{dac90})
and it relies on the metallicity-dependent opacities in cool
stars. The RGB slope is not affected by the reddening, which
is an important advantage in studies of heavily obscured
objects. A critical point is to remove the field contamination
which can affect the RGB slope because it represents a
population with different physical parameters. To achieve this
we used a Monte-Carlo simulation dividing the RGB locus --
spanning the ranges 10.8$\leq$$K$$_S$$\leq$16.8\,mag and
4.3$\leq$$J$$-$$K$$_S$$\leq$5.7\,mag -- on the CMD into bins
with sizes 0.5 and 1.2\,mag along the X and Y axis,
respectively. Then, we randomly removed from each bin of the
cluster CMD as many stars as were present in the corresponding
bin of the field CMD, and determined the cluster RGB slope in
two iterations, removing the 10$\sigma$ outliers. This was
an intentionally selected conservative constraint to ensure that
we do not remove RGB stars or introduce magnitude-related
biases. We repeated this process 300 times and calculated the
average slope and the slope r.m.s.: $-$0.137$\pm$0.012. The
calibration of Ivanov \& Borissova (\cite{iva02a}) yields
[Fe/H]=$-$0.38$\pm$0.14 in the scale of Zinn (as implemented
in Harris \cite{har96}), and [Fe/H]$_{\rm CG}$=$-$0.34$\pm$0.12 in the scale
of Carretta \& Gratton (\cite{car97}).

\begin{figure}[h]
\resizebox{\hsize}{!}{\includegraphics{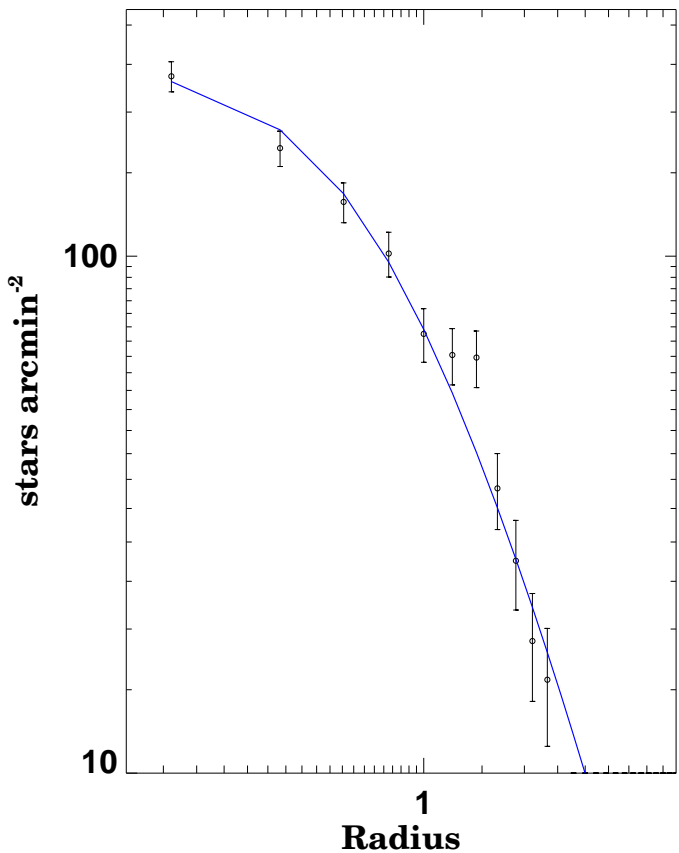}}
\caption{Stellar RPD for the cluster G02 (open circles).
The solid curve represents the best fit King-like profile.
The radius is in arcmin. The X and Y axes are in
logarithmic scale.  
}
\label{fig6}
\end{figure}

Given the uncertainties of both methods, the results are
consistent within the errors and we can only state that G02
is among the most metal rich GC in the Galaxy. We adopt the
average Zinn scale metallicity of [Fe/H]=$-$0.33$\pm$0.14,
where the error is tentatively assigned to the worse error
of the two estimates. The corresponding mean metallicity 
in the scale of Carretta and Gratton is [Fe/H]$_{\rm CG}$=$-$0.16$\pm$0.14. 
In ``The Milky Way Globular 
Cluster Catalog'', 7-th edition of Harris (\cite{har96})
one can find only five clusters with higher metallicity:
Liller\,1, Terzan\,5, NGC\,6528, Palomar\,10, and NGC\,6553.   

\subsection{Structural parameters and position in the Milky Way}

To obtain reliable structural parameters and to separate the 
intrinsic CMD morphology of the cluster from the field we apply 
the statistical decontamination procedure
described in Bonatto \& Bica (\cite{bonatto08}). Typical cell 
dimensions used are $\Delta K$=$0.5$\,mag and $\Delta$($J$$-$$H$)=$\Delta$($J$$-$$K$)=$0.25$\,mag, 
optimal to allow sufficient star-count statistics in the individual cells,
without affecting the morphology of the cluster CMD. 
The comparison field was the same as in the previous sections.

The structural parameters of G02 were determined using the 
iterative star count method of King (\cite{kin62}). The best 
fit to the radial surface brightness profile with a single-mass 
King's model yields a core radius $\rm r_c$=0.70 arcmin (0.9\,pc), 
tidal radius $\rm r_t$=15 arcmin (20\,pc), and central concentration 
c=1.33. The radial density profile (RDP) of the cluster together with 
the best fit King-like profile are presented in Fig.\,\ref{fig6}.
The bins at 1.4 and 1.6 arcmin differ from the profile (in one and three 
standard deviations respectively) and present an excess of the stellar RPD. 
The most probable explanation of this fact is the variable extinction toward
the cluster caused by the clumpy structure of the absorbing matter. 
Nevertheless, the Kolmogorov-Smirnov tests give a very 
satisfactory probability of 0.97 for this profile fit.

The cluster is situated very near to the Galactic plane 
with galactic coordinates X=4.46, Y=1.12 and Z=$-$0.05 in kpc
(defined as in Harris (\cite{har96}), assuming D=4.6 kpc). 
It is located inside the thin disk, at its 
inner edge and the transition region with the bulge.

\section{Summary}

We found that the object No.\,3 from the list of Mercer et al.
(\cite{mer05}) is a new Milky Way globular cluster, and designate
it GLIMPSE-C02 as the second GC discovered from GLIMPSE data.
The analysis based on deep $JHK_S$ images and moderate resolution
K-band spectra of three probable members reveals a
compact, metal rich cluster, with [Fe/H]=$-$0.33$\pm$0.14 in the
scale of Zinn. The cluster is situated near the Galactic plane
behind A$_V$$\sim$24.8$\pm$3\,mag of visual extinction, 
inside the thin disk, at its inner edge and the transition 
region with the bulge. The mean distance estimated from the red 
giant clump stars, tip of the RGB, and isochrone fits 
is D$\sim$4.6$\pm$0.7\,kpc, placing G02 in the zone of 
the ``missing'' globulars.

\begin{acknowledgements}
RK acknowledges support from Proyecto Fondecyt Regular \#\,1080154 
and DIPUV grant No 36/2006 Universidad de Valparaiso, Chile.
Support for JB is provided by Proyecto Fondecyt Regular \#\,1080086, 
Centro de Astrof´ýsica FONDAP No. 15010003, the Chilean
Centro de Excelencia en Astrof´ýsica y Tecnolog´ýas Afines (CATA), 
and DIPUV grant No 36/2006, Universidad de Valparaiso.
This publication makes use of data products from the Two Micron All 
Sky Survey, which is a joint project of the University of Massachusetts 
and the Infrared Processing and Analysis Center/California Institute 
of Technology, funded by the National Aeronautics and Space Administration
and the National Science Foundation. This research has made
use of the SIMBAD database, operated at CDS, Strasbourg, France.
\end{acknowledgements}

\end{document}